\documentclass[a4paper]{jpconf}
\usepackage{graphicx}
\begin{document}
\title{Low-energy reactor neutrino physics with the CONNIE experiment}

\author{Irina Nasteva\footnote{on behalf of the CONNIE collaboration.}} 

\address{Instituto de Física, Universidade Federal do Rio de Janeiro, Av. Athos da Silveira Ramos, 149
CT bloco A, Cidade Universitária, Rio de Janeiro - RJ 21941-972, Brazil}

\ead{Irina.Nasteva@cern.ch}

\begin{abstract}
The Coherent Neutrino-Nucleus Interaction Experiment (CONNIE) uses fully depleted high-resistivity CCDs (charge coupled devices) with the aim of detecting the coherent elastic scattering of reactor antineutrinos off silicon nuclei and probing physics beyond the Standard Model. 
The analysis of the 2016--2018 data allowed us to set an upper limit at 95\% confidence level on the coherent scattering rate, which was used to place stringent constraints on simplified extensions of the Standard Model with light scalar and vector mediators. In 2019, the experiment operated with an improved readout and a lower energy threshold of 50 eV\@. We present the performance of the CONNIE experiment, new results of the analysis of 2019 data, and the recent update of the detector with skipper CCDs. 
\end{abstract}

\section{Coherent elastic neutrino-nucleus scattering}

Although the coherent elastic neutrino-nucleus scattering (CE$\nu$NS) interaction process was predicted by the Standard Model (SM) over 40 years ago~\cite{cenns}, its detection was only made possible recently through the development of very low-threshold detectors~\cite{coherent,coherentAr}. 
Thanks to the coherent enhancement of the cross-section, it is the dominant interaction process at low neutrino energies ($E<50$~MeV) and can be used to probe new physics at this scale. 

Since the coherent scattering rates are predicted with precision by the SM, any deviation could be interpreted a sign of new interactions. 
Several types of non-standard neutrino interactions (NSI) predicted by extensions of the SM  can be explored at low energies using CE$\nu$NS~\cite{Papoulias:2019}. 
Examples include models with light scalar mediators~\cite{Dent:2016, Farzan:2018}, light sterile neutrinos~\cite{Kosmas:2017zbh} and other non-standard neutrino properties, such as millicharge~\cite{Parada:2019gvy}. 
Models in which the neutrino has an anomalous magnetic moment, so that the neutrino-nucleus scattering is mediated by a light boson, predict a significant enhancement of the cross-section at low energies and could result in a several orders of magnitude increase in the event rates~\cite{Miranda:2019, Harnik:2012ni}.
The CE$\nu$NS process is also relevant to supernova energy transport and direct dark matter detection where it is a limiting background. 
Once the detection is well established, the scattering could be used to measure the weak mixing angle~\cite{Canas:2018rng}. 
Additionally, in recent years there has been a growing interest in non-invasive nuclear reactor monitoring using neutrinos~\cite{Cogswell:2016aog}.

\section{The CONNIE experiment}

The experimental challenge of detecting CE$\nu$NS consists in the fact that neutrinos have low energies ($E\sim 1$~MeV for reactor neutrinos), the nuclear recoils that they cause in the interaction have lower energies still (typical recoil energies are of the order of keV), and only a fraction of the recoil energy causes ionization of the detector material. The latter is reflected in the so-called quenching factors that measure the ionization efficiency as a function of nuclear recoil energy. 

The Coherent Neutrino-Nucleus Interaction Experiment (CONNIE)~\cite{connie2015} aims to detect the coherent scattering of reactor antineutrinos off silicon nuclei using fully depleted high-resistivity silicon CCDs (charge coupled devices). In order to achieve the sensitivity needed for measuring the sub-keV ionization due to the recoils of silicon nuclei, it is necessary to lower as much as possible the detection energy threshold and readout noise of the sensors.

The CCD sensors used by CONNIE were developed by LBNL Micro Systems Labs in collaboration with the experiment. 
The sensor consists of a square array with 16 million square pixels of $15\times15$~$\mu$m$^2$ pitch each and 657~$\mu$m thickness. 
The 14 CCDs are installed in a copper box, which is kept in a copper vacuum vessel.
In order to reduce the thermally-generated dark current, the sensors are cooled to temperatures below 100~K and operate in a vacuum of $10^{-7}$~torr.
The detector is surrounded by passive shielding, formed by 15~cm of lead to absorb photons, sandwiched between two 30-cm layers of high-density polyethilene to stop cosmogenic neutrons.

CONNIE is located about 30 m from the core of the 3.8 GW Angra 2 nuclear reactor in Rio de Janeiro, Brazil. It is positioned in a shipping container lab at sea level just outside the reactor dome, where the neutrino flux density is estimated to be $7.8\times 10^{12}\ \bar{\nu}$s$^{-1}$cm$^{-2}$. 
The detector was first installed in 2014 with engineering-grade CCDs, which were employed to commission its operation and characterise backgrounds~\cite{connie2016}. 
Since the installation of scientific CCDs in 2016, the experiment has been taking data continuously with small interruptions for servicing.

\section{Results from the 2016--2018 run}

\begin{figure}[tb]
\includegraphics[width=20pc]{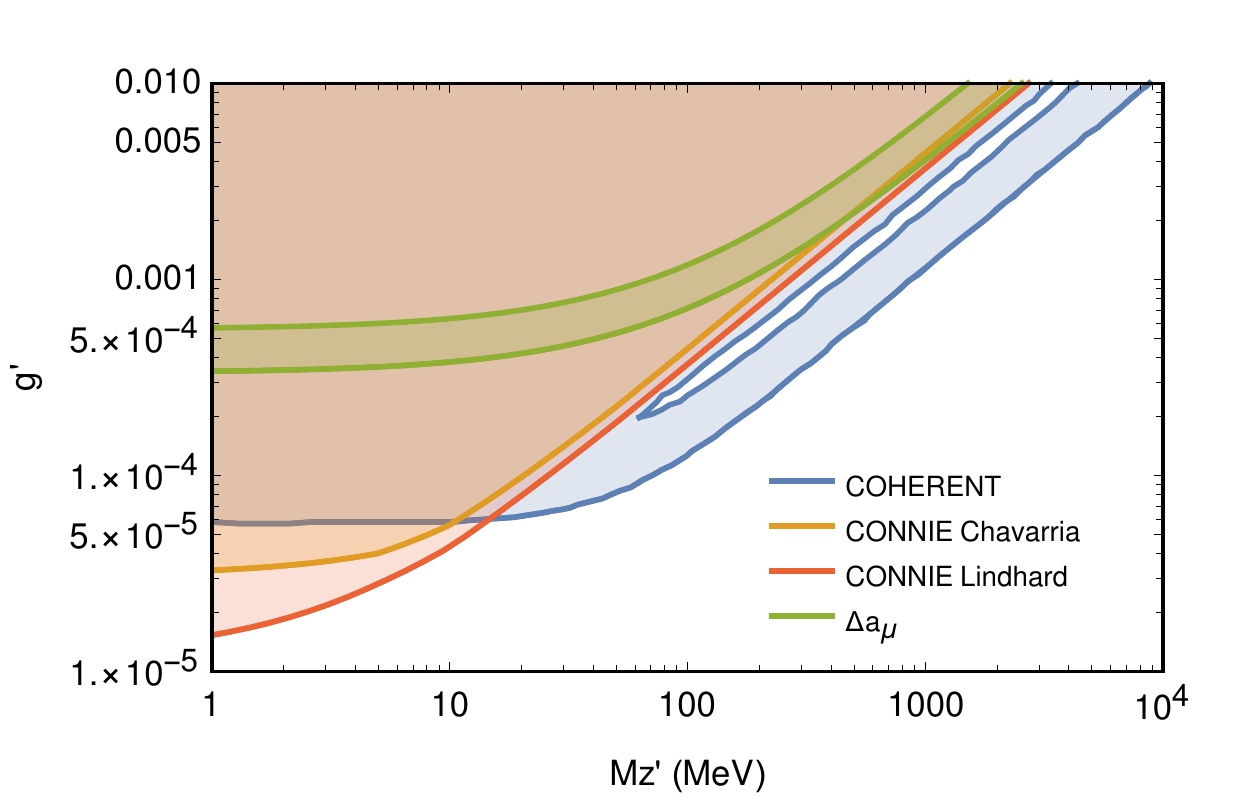}
\includegraphics[width=19pc]{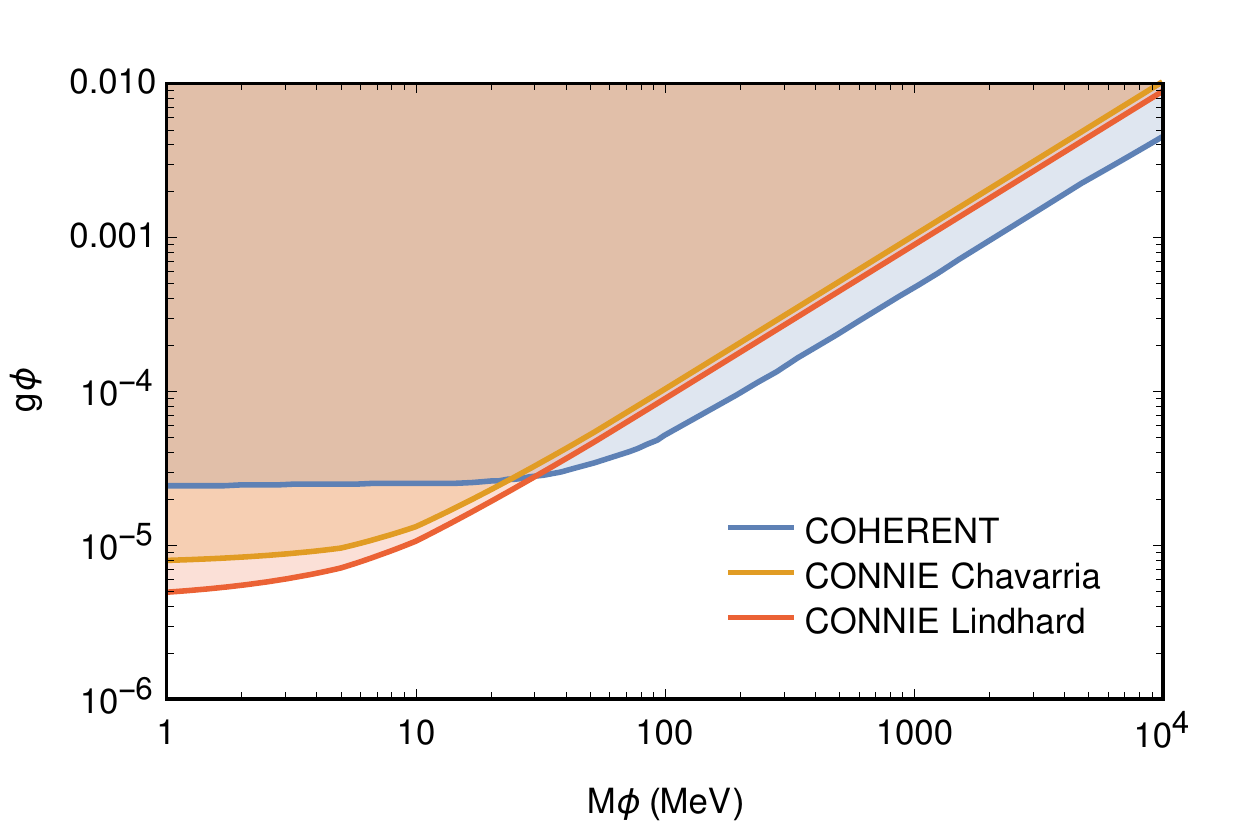}
\caption{\label{fig:mediators}CONNIE exclusion limits in the phase space of a simplified model with a light vector mediator $Z'$ (left) and with a light scalar mediator $\phi$ (right)~\cite{connieLM}, obtained from CONNIE 2016--2018 data with the Lindhard~\cite{lindhard} and Chavarria~\cite{chicago_qf} quenching factors.}
\end{figure}

The data are recorded as images of three-hour exposure each, which are then processed and the physical charge deposition events extracted. 
The energy (or gain) calibration is performed in situ by using copper fluorescence peaks, while the depth of the interaction is determined from the size of the lateral charge diffusion, by applying a depth-diffusion curve measured from a sample of straight muon tracks. 
The gain stability, as well as readout noise and background levels from cosmic muons and copper radioactivity are monitored continuously.

The total accumulated exposure of the 2016--2018 data sample corresponds to 8 good-performance CCDs (47.6 g active mass) and 3.7 kg-days: 2.1 kg-days taken with the reactor on and 1.6 kg-days with the reactor off. 
Events are selected to contain energy above a threshold, 
corresponding to roughly four times the standard deviation of the readout noise.  
A statistical test is used to separate neutrino-like events from spurious ones from on-chip noise sources at low energies, based on the likelihood of the charge deposition of an event to follow the probability density function of the Gaussian readout noise.
The energy spectra are studied with the reactor on an off, and show good agreement without any excess at low energies. 
The difference between the two is used to obtain an upper limit at 95\% confidence level (CL) on the coherent neutrino scattering rate~\cite{connie2019}, which corresponds to about 40 times the SM prediction at recoil energies down to 1~keV (0.1~keV measured energy).
These results constitute the first search for CE$\nu$NS at a nuclear reactor reaching such low energies.

The upper limit in the lowest-energy bin is used to constrain two simplified SM extension models, in which non-standard neutrino interactions are mediated by a light vector $Z'$ and a light scalar mediator $\phi$. 
Figure~\ref{fig:mediators} shows the resulting exclusion limits in the phase space of the mediador mass and coupling for the two models. 
These constraints~\cite{connieLM} represent the best limits for these simplified models among the CE$\nu$NS experiments for a light vector mediator with mass $M_{Z'} <10$~MeV, and for a light scalar mediator with mass $M_{\phi}<30$~MeV\@, as well as the first competitive beyond the SM constraints from CE$\nu$NS
at reactors.

\section{Improvements and analysis of 2019 data}

\begin{figure}[tb]
\centering
\includegraphics[width=28pc]{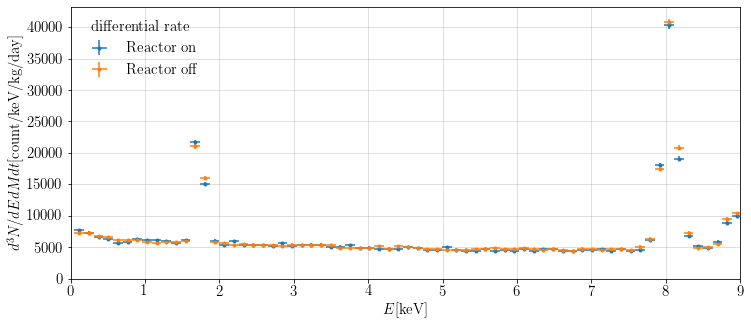}
\includegraphics[width=28pc]{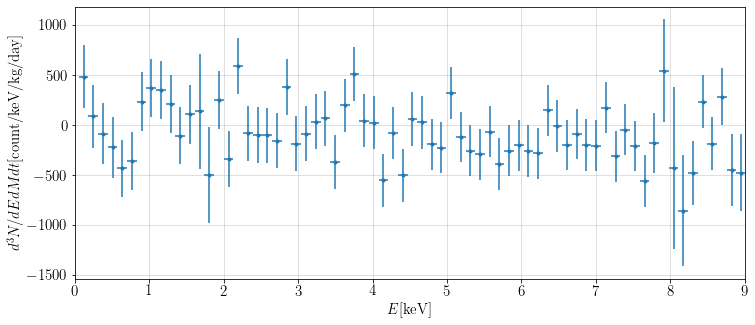}
\caption{\label{connie2019_onoff}CONNIE total event rates measured with the reactor on and off (top) and difference between the reactor-on and reactor-off event rates (bottom) in bins of measured energy and averaged over CCDs, obtained from 2019 data~\cite{connie2021}.}
\end{figure}

\begin{figure}[h]
\centering
\includegraphics[width=28pc]{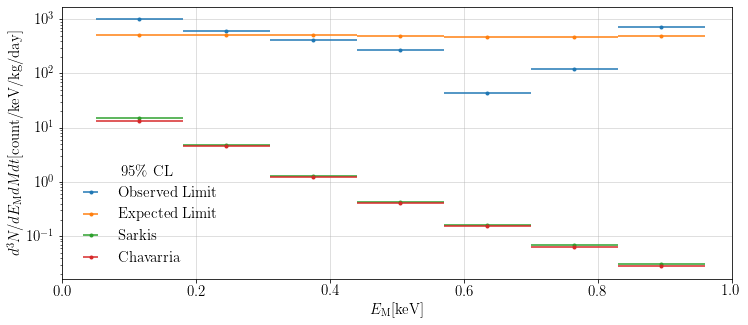}
\caption{\label{connie2019_limit}Upper limits on the coherent neutrino interaction rate at 95\% CL from CONNIE 2019 data~\cite{connie2021}. The observed limit (blue) and expected limit (orange) in bins of measured energy are compared with the standard model rate predictions, calculated with quenching factors from Sarkis~\cite{QF} (green) and Chavarria~\cite{chicago_qf} (red).}
\end{figure}

In 2019 the experiment operated with an improved readout mode, by applying a hardware binning to the readout stage of $5\times 1$ pixels in the vertical direction, which led to reduced levels of readout noise and allowed to lower the detection threshold to 50 eV~\cite{connie2021}. 
The exposure times were reduced to 1 hour per image, resulting in lower spurious charge accumulation. 
In addition, improvements were applied in the data analysis, such as new gain and depth calibrations and background characterisation. 
On-chip noise sources were quantified in the reactor-off data, and studies of spurious events that can mimic a neutrino signal at very low energies were conducted and helped reduce these dominant backgrounds and improve the sensitivity to neutrinos. 

The 2019 CONNIE data sample corresponds to 31.85 days of reactor-on and 28.25 days of reactor-off operation with 8 CCDs, whose total fiducial mass after geometric and event size cuts is 36.24~g, giving an experiment exposure of 2.2 kg-days.
The analysis was developed in a blind manner based on reactor-off data. 
After unblinding, the reactor-on and reactor-off spectra were compared, as shown in Fig.~\ref{connie2019_onoff}, and were found to be consistent with each other and the difference statistically consistent with zero. 
The background rate at low energies is reduced with respect to the previous study, thanks to the binned readout and new analysis techniques.

From the difference in the lowest-energy bins, upper limits at 95\% CL are placed on the coherent neutrino scattering rates as a function of measured energy.
The observed and expected limits, together with the rates predicted by the SM using two different quenching factor models, are shown in Fig.~\ref{connie2019_limit}. 
Using the newer Sarkis quenching factor~\cite{QF} which makes use of binding energy to extend to lower recoil energies, the resulting expected (observed) limit stands at 34~(66) times the SM prediction in the $50-180$~eV measured energy range. 
The sensitivity in 2019 data is greatly enhanced with respect to the previous results, reaching energies down to 50~eV.

\section{Update with skipper CCDs and perspectives}

\begin{figure}[tb]
\centering
\includegraphics[width=15pc]{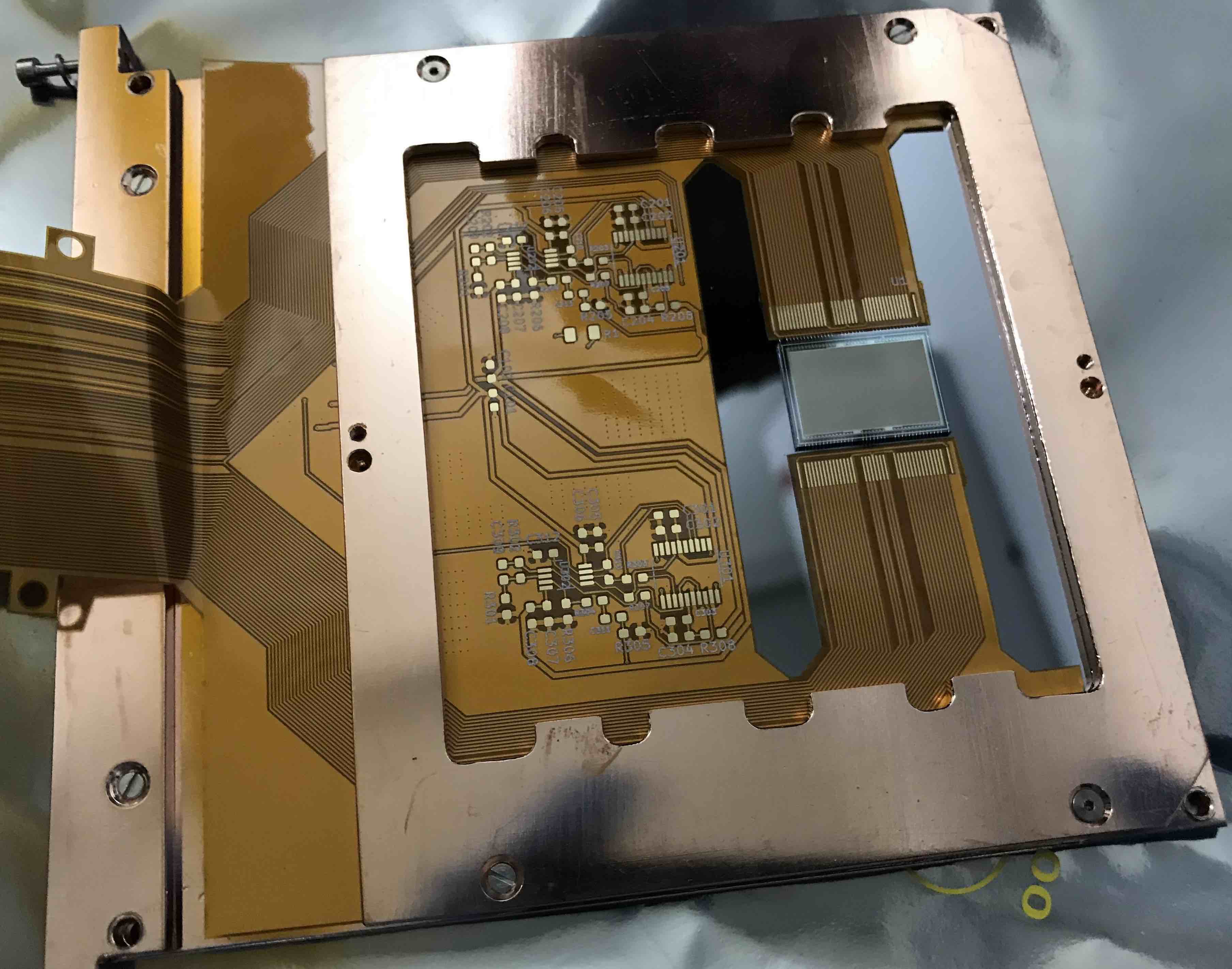}
\includegraphics[width=22pc]{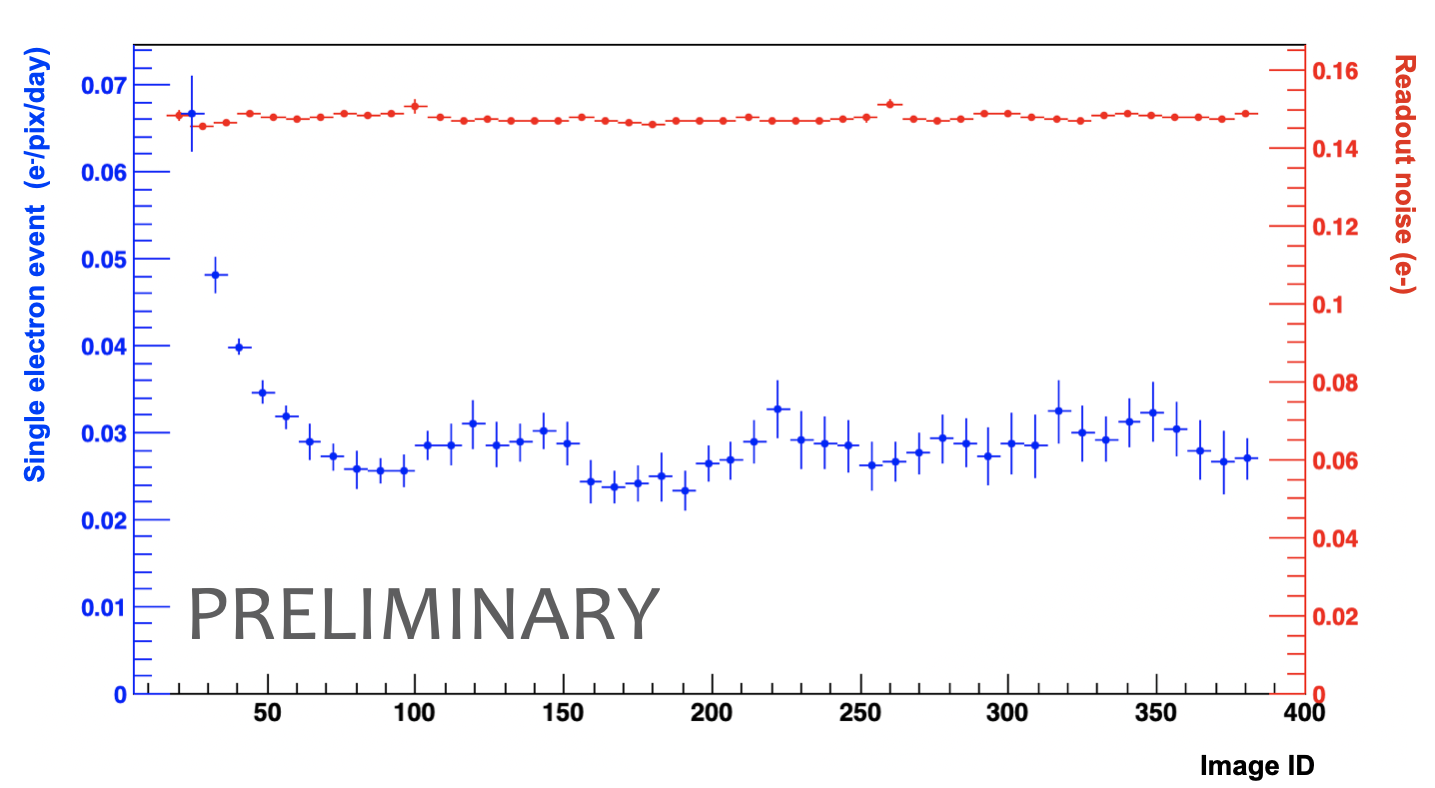}
\caption{\label{connie_skipper}A skipper CCD that was installed in the CONNIE detector in 2021 (left) and levels of the single electron events and readout noise measured with time (right).}
\end{figure}

In order to further lower the detection threshold and improve sensitivity, CONNIE plans to employ the recently developed skipper CCD sensors~\cite{skipper}.  
Skipper CCDs operate with multiple non-destructive sampling of the pixel charge, leading to very low readout noise down to sub-electron levels, which allows to count the number of individual electrons in each pixel.

In July 2021, the CONNIE detector was updated with two skipper CCD sensors (Fig.~\ref{connie_skipper}) of $1022 \times 682$ pixels each, 675~$\mu$m thickness and a total mass of 0.4~g.  
The sensors were installed together with new Low Threshold Acquisition readout electronics~\cite{lta} and a new dedicated vacuum interface signal transfer board. 
After a short period of operation without shielding, the setup is currently taking data with partial shielding with the initial aim of optimising the readout mode and characterising the skipper CCD performance. 
Figure~\ref{connie_skipper} illustrates the preliminary stable and very low values of spurious charge or single electron events ($\sim 0.03$~e$^-$/pix/day) and readout noise ($\sim 0.15$~electrons) achieved so far.

This characterisation of skipper CCD sensors and sea-level background will help prepare for future larger-mass skipper CCD experiments.
The next steps for CONNIE include completing the detector with more skipper CCDs to increase its mass and collecting data with the full shielding in order to achieve the projected sensitivity.

\ack
This work was partially supported by the Brazilian funding agencies FAPERJ and CPNq.

\section*{References}

\bibliography{connie_bibliography}

\providecommand{\newblock}{}
\begin{thebibliography}{10}
\expandafter\ifx\csname url\endcsname\relax
  \def\url#1{{\tt #1}}\fi
\expandafter\ifx\csname urlprefix\endcsname\relax\def\urlprefix{URL }\fi
\providecommand{\eprint}[2][]{\url{#2}}

\bibitem{cenns}
Freedman D~Z 1974 {\em Phys. Rev. D\/} {\bf 9} 1389--1392

\bibitem{coherent}
Akimov D {\em et~al.\/} (COHERENT) 2017 {\em Science\/} {\bf 357} 1123--1126
  (\textit{Preprint} \eprint{1708.01294})

\bibitem{coherentAr}
Akimov D {\em et~al.\/} (COHERENT) 2021 {\em Phys. Rev. Lett.\/} {\bf 126}
  012002 (\textit{Preprint} \eprint{2003.10630})

\bibitem{Papoulias:2019}
Papoulias D~K {\em et~al.\/} 2019 {\em Front. in Phys.\/} {\bf 7} 191
  (\textit{Preprint} \eprint{1911.00916})

\bibitem{Dent:2016}
Dent J~B, Dutta B, Liao S, Newstead J~L, Strigari L~E and Walker J~W 2017 {\em
  Phys. Rev. D\/} {\bf 96} 095007 (\textit{Preprint} \eprint{1612.06350})

\bibitem{Farzan:2018}
Farzan Y {\em et~al.\/} 2018 {\em JHEP\/} {\bf 05} 066 (\textit{Preprint}
  \eprint{1802.05171})

\bibitem{Kosmas:2017zbh}
Kosmas T~S, Papoulias D~K, Tortola M and Valle J~W~F 2017 {\em Phys. Rev. D\/}
  {\bf 96} 063013 (\textit{Preprint} \eprint{1703.00054})

\bibitem{Parada:2019gvy}
Parada A 2020 {\em Adv. High Energy Phys.\/} {\bf 2020} 5908904
  (\textit{Preprint} \eprint{1907.04942})

\bibitem{Miranda:2019}
Miranda O~G {\em et~al.\/} 2019 {\em JHEP\/} {\bf 07} 103 (\textit{Preprint}
  \eprint{1905.03750})

\bibitem{Harnik:2012ni}
Harnik R, Kopp J and Machado P~A~N 2012 {\em JCAP\/} {\bf 07} 026
  (\textit{Preprint} \eprint{1202.6073})

\bibitem{Canas:2018rng}
Ca\~nas B~C, Garc\'es E~A, Miranda O~G and Parada A 2018 {\em Phys. Lett. B\/}
  {\bf 784} 159--162 (\textit{Preprint} \eprint{1806.01310})

\bibitem{Cogswell:2016aog}
Cogswell B~K and Huber P 2016 {\em Science \& Global Security\/} {\bf 24}
  114--130

\bibitem{connie2015}
Fernandez~Moroni G {\em et~al.\/} 2015 {\em Phys. Rev. D\/} {\bf 91} 072001
  (\textit{Preprint} \eprint{1405.5761})

\bibitem{connie2016}
Aguilar-Arevalo A {\em et~al.\/} (CONNIE) 2016 {\em JINST\/} {\bf 11} P07024
  (\textit{Preprint} \eprint{1604.01343})

\bibitem{connieLM}
Aguilar-Arevalo A {\em et~al.\/} (CONNIE) 2020 {\em JHEP\/} {\bf 04} 054
  (\textit{Preprint} \eprint{1910.04951})

\bibitem{lindhard}
James F~{Ziegler} J P~{Biersack} U~L 1985 {\em The Stopping and Range of Ions
  in Solids\/} (Pergamon)

\bibitem{chicago_qf}
Chavarria A~E {\em et~al.\/} 2016 {\em Phys. Rev. D\/} {\bf 94} 082007
  (\textit{Preprint} \eprint{1608.00957})

\bibitem{connie2019}
Aguilar-Arevalo A {\em et~al.\/} (CONNIE) 2019 {\em Phys. Rev. D\/} {\bf 100}
  092005 (\textit{Preprint} \eprint{1906.02200})

\bibitem{connie2021}
Aguilar-Arevalo A {\em et~al.\/} (CONNIE) 2021  (\textit{Preprint}
  \eprint{2110.13033})

\bibitem{QF}
Sarkis Y, Aguilar-Arevalo A and D'Olivo J~C 2020 {\em Phys. Rev. D\/} {\bf 101}
  102001 (\textit{Preprint} \eprint{2001.06503})

\bibitem{skipper}
Tiffenberg J {\em et~al.\/} (SENSEI) 2017 {\em Phys. Rev. Lett.\/} {\bf 119}
  131802 (\textit{Preprint} \eprint{1706.00028})

\bibitem{lta}
Cancelo G~I {\em et~al.\/} 2021 {\em J. Astron. Telesc. Instrum. Syst.\/} {\bf
  7} 015001 (\textit{Preprint} \eprint{2004.07599})

\end{thebibliography}

\end{document}